\documentclass[aps,prb,twocolumn,showpacs,superscriptaddress]{revtex4-1}\begin{tiny}\end{tiny}
\usepackage{longtable}
\usepackage{amsfonts}
\usepackage{graphicx}
\usepackage{epsfig}

\usepackage{setspace}

\begin{document}

\title{Improved Callaway model for Lattice Thermal Conductivity}

\author{Philip B. Allen}
 \email{philip.allen@stonybrook.edu}
 \affiliation{Physics and Astronomy Department, SUNY Stony Brook University, NY 11794-3800, USA}

\date{\today}


\begin{abstract}
In developing the phonon quasiparticle picture, Peierls discovered that, in a perfect crystal, without anharmonic Umklapp ($U$) 
events, a current-carrying
distribution can never relax to a zero-current distribution.  Callaway introduced a simplified approximate model
version of the Peierls-Boltzmann equation, retaining its ability to deal separately with Normal ($N$) and $U$ events.  This paper
clarifies and improves the Callaway model, and shows that Callaway underestimated the suppression of $N$-processes
in relaxing thermal current.  The new result should improve computations of thermal conductivity
from relaxation-time studies.
\end{abstract}

\maketitle


\section{Introduction}

Debye \cite{Debye} was the first to realize that a perfect harmonic crystal is a perfect heat conductor.  In insulators, heat is carried
by propagating lattice-vibrational normal modes.  Quantum theory simplifies by identifying
these modes as particles.  Anharmonic interactions permit
a phonon with wave-vector $\vec{Q}$ to interact with other phonons $\vec{Q}^\prime$ and $\vec{Q}^{\prime\prime}$ (either by decay
into two, or absorption of one and emission of the other.)  These events cause thermalization and resistance to current flow.
The Peierls-Boltzmann Equation (PBE) \cite{Peierls} puts this on a firm footing.  Peierls made the important observation that
wave-vector-conserving (``Normal" or ``$N$") events ($\vec{Q}=\vec{Q}^\prime \pm \vec{Q}^{\prime\prime}$) 
cannot by themselves cause a current-carrying state
(with non-zero total wave-vector) to decay to a zero current state, but that Umklapp (``$U$") events 
($\vec{Q}=\vec{Q}^\prime \pm \vec{Q}^{\prime\prime} + \vec{G}$, where $\vec{G}$ is a reciprocal lattice vector), can relax
the current to zero.

Solving the PBE is still not an easy task, but modern advances make it possible.  In particular, ``{\it ab initio}" computation of harmonic
normal modes is now very successful \cite{Baroni}; similar techniques give quite reliable anharmonic forces \cite{anh}; thus
full solutions from purely theoretical input are now being done with good success \cite{Broido}.  This does not diminish
the need for simplified approximate models, to enable us to think about the physics of the process, and perhaps to invent reliable
approximate treatments that avoid the full solution.  Such a model was introduced by Callaway \cite{Callaway} in 1959.  This model 
had a big influence on the field, although the model is rarely used in detail.  Re-thinking Callaway's model has allowed me to
improve it, correcting and simplifying the solution.  Then using a Debye-type phonon model, the relative role of $N$
and $U$ processes is re-analyzed.

\section{Peierls-Boltzmann Equation}

The PBE is
\begin{equation}
\frac{dN_Q}{dt} = \frac{\partial N_Q}{\partial t}  -\vec{v}_Q \cdot \frac{\partial N_Q}{\partial \vec{r}} + \left[\frac{\partial N_Q}{\partial t}\right]_{\rm collision},
\label{eq:PBE}
\end{equation}
where $Q$ is short for $(\vec{Q},n)$ -- both wave-vector and the ``branch indices'' $n$ needed to specify a
propagating vibrational normal mode.  The phonon group velocity is $\vec{v}_Q = \partial\omega_Q/\partial\vec{Q}$, where 
$\omega_Q$ is the frequency.  The term ``quasiparticle''  denotes a propagating vibrational normal mode.  Disorder 
and anharmonic interactions must not be so strong as to make the mean free path $\ell$ as
short as a wavelength.  The quasiparticle picture breaks down if the wavevector uncertainty
caused by scattering is too large; $\omega_Q$ and $\vec{v}_Q$ are correspondingly poorly defined.
If the heat-carrying excitations are not quasiparticles, then a theory more complicated
than the PBE is needed.  The Ioffe-Regel criterion\cite{Ioffe}, which says no currents can flow
if quasiparticles are destroyed ($\ell < \lambda$) is not correct.  Better theories
are sometimes available \cite{Parshin,Allen}.   According to the PBE, the
distribution $N_Q$ may vary in space, and, if not driven, will relax under collisions to a local
equilibrium Bose-Einstein distribution $n_Q(T(\vec{r}))$.  The collision term is
a complicated non-linear sum over other phonon states $N_{Q^{\prime}}$.   The collisions
conserve local energy,
\begin{equation}
E = \sum_Q \hbar\omega_Q \left( N_Q(\vec{r}) + \frac{1}{2} \right).
\label{eq:U}
\end{equation}
In the absence of anharmonic $U$ events or other momentum-non-conserving processes (e.g. disorder), the total local wavevector 
\begin{equation}
\vec{P}(\vec{r}) = \sum_Q \vec{Q} N_Q (\vec{r})
\label{eq:P}
\end{equation}
is also conserved under collisions.

The full Peierls-Boltzmann equation has an important property, namely, generating the Boltzmann ``H-theorem.''
\cite{Pauli}   As explained, {\it e.g.} by Landau and Lifshitz \cite{Landau}, counting the multiplicity of states gives a quasiparticle entropy 
\begin{equation}
S = k_B\sum_Q[(N_Q +1)\ln(N_Q +1)-N_Q \ln N_Q].
\label{eq:S}
\end{equation}
When this is maximized, subject to the constraint of constant $E$, Eq.(\ref{eq:U}), the result is $N_Q \rightarrow n_Q$.
Even though entropy is strictly not defined except in equilibrium, nevertheless, Eq.(\ref{eq:S}) qualifies as a non-equilibrium
local quasiparticle entropy.  When the PBE is used to compute the rate of change $dS/dt$ of Eq.(\ref{eq:S}),  
one can show that $(dS/dt)_{\rm collision} \ge 0$. 
The distribution $N_Q$ that is stationary under collisions is the one that maximizes $S$ under the relevant constraints.
If the sole type of collision is anharmonic phonon scattering with only $N$ processes, then the relevant object to maximize is
\begin{equation}
S/k_B - \beta E - \vec{\Lambda} \cdot \vec{P},
\label{eq:SLagrange}
\end{equation}
where $\beta$ and $\vec{\Lambda}$ are Lagrange multipliers.  The maximum occurs when
$N_Q$ evolves to a ``flowing equilibrium'' $n_Q^\ast$,
\begin{equation}
N_Q \rightarrow n_Q^\ast = \left[ e^{\hbar\omega_Q/k_B T +\vec{\Lambda} \cdot \vec{Q}}-1 \right]^{-1}.
\label{eq:dispBE}
\end{equation}
The Lagrange multiplier $\beta$ has been identified as $1/k_B T$, with a local temperature $T(\vec{r})$,
but the Lagrange multiplier $\vec{\Lambda}$ has still to be identified.

Now consider heat transport under an impressed temperature gradient.
The steady state distribution obeys $dN_Q/dt = \partial N_Q /\partial t = 0$.  
The distribution function
relaxes toward a local equilibrium ($T$ may vary spatially), with a small deviation $N_Q \rightarrow n_Q +\Phi_Q$.
The aim is to find, to first order in $\vec{\nabla} T$, the heat current, defined as
\begin{equation}
\vec{j}=(1/\Omega)\sum_Q \hbar\omega_Q \vec{v}_Q \Phi_Q \equiv -{\sf \kappa} \cdot \vec{\nabla}T
\label{eq:j}
\end{equation}
The spatial gradient term in the PBE Eq.(\ref{eq:PBE}) can be linearized in the thermal gradient; $\vec{v}_Q \cdot \vec{\nabla}N_Q$ 
becomes $(\partial n_Q / \partial T) \vec{v}_Q\cdot\vec{\nabla}T$.  
It common to approximate the collision term by a simplified linear and local-in-$Q$ approximation.
The steady-state, linear in $\vec{\nabla}T$, PBE, and the corresponding thermal conductivity, become
\begin{equation}
0 \approx -\vec{v}_Q \cdot \vec{\nabla}T \ \partial n_Q/\partial T -\Phi_Q/\tau_Q,
\label{eq:relaxPB}
\end{equation}
\begin{equation}
\kappa_{\rm RTA} =\frac{1}{\Omega}\sum_Q \hbar\omega_Q v_{Qx}^2 \tau_Q \  \partial n_Q/\partial T.
\label{eq:kappa-0}
\end{equation}
This is the relaxation-time approximation (RTA), and $\tau_Q$ is the phonon relaxation time.  
It is worth noting that, although this represents a serious
approximation to the full PBE, nevertheless, the exact solution, if available, can always be put in the form of Eq.(\ref{eq:kappa-0}), with
a suitably redefined relaxation time.  The idea is
that the exact distribution function, $n_Q + \Phi_Q$, found by solving the linearized integral equation, can be written,
in linear approximation, as
$\Phi_Q^{\rm exact} \equiv  -\tau_Q^{\rm exact}\vec{v}_Q \cdot \vec{\nabla}T \ \partial n_Q/\partial T$.
This defines a quantity $\tau_Q^{\rm exact}$ which can be interpreted as the time for current relaxation
in the channel $Q$.  It is perhaps not very different from, but is surely not the same as, the ordinary
``quasiparticle'' relaxation time, defined using thermal Green's function theory via the self-energy ($\Sigma$),
$1/\tau_Q^{\rm QP} \equiv -2\rm{Im}\Sigma(Q,\omega+i\eta)$.

\section{approximate treatment of $N$ {\it versus} $U$}

The question is, how to find an approximate $\tau_Q$ that will give an accurate thermal
conductivity, without the full labor of solving the PBE?  The choice $\tau_Q^{\rm QP}$ has some
advantages, since it is a well-defined object, measureable by neutron or x-ray scattering, and not
overwhelming to compute by modern methods.  It is also an object of interest in Peierls-Boltzmann
theory.  If all phonons $Q^\prime$ are forced to be in equilibrium ($N_{Q^\prime}=n_{Q^\prime}$) except when $Q^\prime$ equals $Q$,
then $(dN_Q/dt)_{\rm collision}$ becomes $ -(N_Q - n_Q)/\tau_Q^{\rm QP}$.
The PBE form for $\tau_Q^{\rm QP}$ agrees with the Green's function 
result in the usual anharmonic perturbation theory.  The
RTA consists of using $\tau_Q^{\rm QP}$ as the $\tau_Q$ in
Eq.(\ref{eq:relaxPB}).  This underestimates the thermal conductivity.  When $N_Q \ne n_Q$, all collisions involving mode $Q$
help relax the quasiparticle population of state $Q$; however, there are $N$ processes which contribute to $\tau_Q^{\rm QP}$ but
cannot be fully active in relaxing the current.  They don't fully contribute to $\tau_Q^{\rm exact}$.
This is where the Callaway model \cite {Callaway} comes in.

Callaway's idea is to write $(\partial N_Q/\partial t)_{\rm collision}$ in  two parts, as $-(N_Q-n_Q)/\tau_Q^U -(N_Q-n_Q^\ast)/\tau_Q^N$.
The collective relaxation rate, 
\begin{equation}
1/\tau_Q^c = 1/\tau_Q^U + 1/\tau_Q^N
\label{eq:tauc}
\end{equation}
is just the total quasiparticle relaxation rate \cite{footnote}.
The part denoted $1/\tau_Q^N$, arising from anharmonic $N$-processes, leaves the total crystal momentum unchanged. 
Only the ``$U$'' part $1/\tau_Q^U$ can relax to the final zero-current equilibrium.  The part $1/\tau_Q^N$ relaxes the 
distribution to the flowing equilibrium $n_Q^\ast$, Eq.(\ref{eq:dispBE}).  When other mechanisms of phonon relaxation, 
such as disorder, are present, they also destroy crystal momentum conservation, and are grouped with the $U$ terms.

The deviation $\Phi_Q = N_Q - n_Q$ (Eq. \ref{eq:j})
determines the current.  Deviation from the flowing equilibrium can be written by Taylor expansion as
$N_Q - n_Q^\ast = \Phi_Q + (k_B T^2/\hbar\omega_Q)(\partial n_Q/\partial T) \vec{\Lambda}\cdot\vec{Q}$.
The Callaway-modified RTA therefore gives the distribution function as
\begin{equation}
\Phi_Q = -\tau_Q^c \vec{v}_Q \cdot \vec{\nabla}T \frac{\partial n_Q}{\partial T} - \frac{\tau_Q^c}{\tau_Q^N}\frac{k_B T^2}
{\hbar\omega_Q} \vec{\Lambda}\cdot\vec{Q} \frac{\partial n_Q}{\partial T}.
\label{eq:Phi}
\end{equation}
The Lagrange multiplier $\vec{\Lambda}$
is not yet determined.  This is where my answer deviates a bit from Callaway's.

The total crystal momentum $\vec{P}$ (Eq. \ref{eq:P}) should be the same for both the actual distribution 
$N_Q$ and the flowing equilibrium
distribution $n_Q^\ast$ that $N$ processes drive $N_Q$ towards.  This means 
\begin{equation}
\sum_Q \vec{Q}(N_Q -n_Q^\ast) = 0=\sum_Q \vec{Q}(\Phi_Q + n_Q - n_Q^\ast)
\label{eq:constraint}
\end{equation}
Taylor expanding gives
\begin{equation}
n_Q - n_Q^\ast=n_Q(n_Q+1)\vec{\Lambda}\cdot\vec{Q}
=\frac{k_B T^2}{\hbar\omega_Q}\frac{\partial n_Q}{\partial T} \vec{\Lambda}\cdot\vec{Q}.
\label{eq:nstar}
\end{equation}
Inserting Eqs.(\ref{eq:Phi},\ref{eq:nstar}) into Eq.(\ref{eq:constraint}) gives an equation for
the Lagrange multiplier $\vec{\Lambda}$,
\begin{equation}
\sum_Q \tau_Q^c  (\vec{v}_Q \cdot \vec{\nabla}T)  \vec{Q} \frac{\partial n_Q}{\partial T} = \sum_Q
\frac{\tau_Q^c}{\tau_Q^U} \frac{k_B T^2}{\hbar\omega_Q} (\vec{\Lambda} \cdot \vec{Q}) \vec{Q} \frac{\partial n_Q}{\partial T}
\label{eq:constr}
\end{equation}
This replaces Eq.(14) of Callaway's paper \cite{Callaway}, which is equivalent except for an extra factor of $1/\tau_Q^N$ inside the
sums on both sides of the equation.  Why does Callaway have a different formula fixing $\vec{\Lambda}$?  Callaway uses
the constraint that the time rate of change of $\vec{P}$ from $N$ processes must vanish.  This is surely an equally exact
statement, but, in order to implement it, Callaway makes an {\bf additional} use of 
the relaxation time model.  This gives an extra factor of $1/\tau_Q^N$
inside both $Q$-sums in Eq.(\ref{eq:constraint}).   The model is inexact, and leads to a difference from Eq.(\ref{eq:constr}), which
made no such additional approximation.  Insofar as Callaway's Eq.(14) differs from Eq.(\ref{eq:constr}), Callaway's method is wrong.

The argument simplifies by assuming cubic symmetry, or else a thermal gradient along a symmetry axis (denoted $x$) 
of an orthorhombic crystal.  Then only $\Lambda_x$ is needed.  Its value cancels from Eq.(\ref{eq:Phi}) when the second
term is multiplied by the left-hand side, and divided by the right-hand side, of Eq.(\ref{eq:constr}).  When the resulting equation
for $\Phi_Q$ is substituted into Eq.(\ref{eq:j}), a formula for the thermal conductivity results,
\begin{equation}
\kappa_{xx}=\kappa_c + \lambda_1 \lambda_2/\lambda_3.
\label{eq:kappaUN}
\end{equation}
 The leading term, $\kappa_c$, is just the usual relaxation time formula,
\begin{equation}
\kappa_c = \frac{1}{\Omega}\sum_Q \hbar\omega_Q v_{Qx}^2 \tau_Q^c \  \partial n_Q/\partial T
\label{eq:kappa-1}
\end{equation}
where $1/\tau_Q^c$ is the usual quasiparticle relaxation rate, $1/\tau_Q^U + 1/\tau_Q^N$, containing
both $N$ and $U$ processes.  The correction factors are
\begin{eqnarray}
\lambda_1 &=& \frac{1}{\Omega}\sum_Q  v_{Qx} Q_x  \tau_Q^c \  \partial n_Q/\partial T
\label{eq:lambda1} \\
\lambda_2 &=& \frac{1}{\Omega}\sum_Q  v_{Qx} Q_x (\tau_Q^c /\tau_Q^N )  \partial n_Q/\partial T
\label{eq:lambda2} \\
\lambda_3 &=& \frac{1}{\Omega}\sum_Q  (Q_x^2 /\hbar\omega_Q) (\tau_Q^c /\tau_Q^U )  \partial n_Q/\partial T
\label{eq:lambda3}
\end{eqnarray}
Eqs.(\ref{eq:kappaUN}-\ref{eq:lambda3}) are the new result of this paper.  They are an approximate procedure, based
on the Callaway model, which better contains the different roles of $N$ and $U$ scattering events.  Callaway's
answer is similar, except that $\tau_Q^c$ is replaced by the ratio $\tau_Q^c/\tau_Q^N$ in both $\lambda_1$ 
(Eq.\ref{eq:lambda1}) and $\lambda_3$ (Eq.\ref{eq:lambda3}).  

To see the consequence of this modification, the Debye model is appropriate.  
It assumes three branches $\omega_Q = vQ$, with $v$ a constant sound velocity, the same
for simplicity, for all three branches.  The dispersion
relations are spherically symmetric, and the Brillouin zone approximated by a sphere
of radius $Q_D$ with maximum frequency $\omega_D=vQ_D$. 
The Debye density of states is $\mathcal{D}(\omega)=(9N/V)\omega^2/\omega_D^3$.
The specific heat is 
\begin{equation}
C(T)=\int_0^{\omega_D} d\omega \hbar\omega \frac{\partial n(\omega)}{\partial T}\mathcal{D}(\omega).
\label{eq:CD}
\end{equation}
In the same spirit, one assumes scattering rates $1/\tau_Q^N$ and $1/\tau_Q^U$ to depend only
on $\omega_Q$ and $T$, that is, the only $Q$-dependence comes through $\omega_Q$.
Furthermore, it is common to assume that the resulting function factorizes into
a power of frequency $\omega$ times a function of $T$, 
\begin{equation}
1/\tau_Q^\alpha \rightarrow 1/\tau_\alpha(\omega_Q,T) = \gamma_\alpha(T)\times(\omega_Q/\omega_D)^{p_\alpha}.
\label{eq:taupower}
\end{equation}
If we look only at ratios, it will not be necessary to choose a $T$-dependence of $\gamma_N$.  For
$\gamma_U$'s, at low $T$,
the needed large $Q$ thermal phonon is thermally suppressed by a factor often written as 
$\gamma_U = \gamma_N \times A \exp(-\Theta_D/aT)$, where $A$ is a constant, independent of $T$, and
$\Theta_D$ is the Debye temperature, $k_B \Theta_D = \hbar \omega_D$. 
The adjustable parameter $a$ is often set to $3$.

If is convenient to define a frequency average $\overline{f}$ of a frequency-dependent function $f(\omega)$ as
\begin{equation}
\overline{f}(T)=\frac{1}{C(T)}\int_0^{\omega_D} d\omega \hbar\omega \frac{\partial n(\omega)}{\partial T}\mathcal{D}(\omega)f(\omega).
\label{eq:fbar}
\end{equation}
Then in the Debye model, the answers Eqs.(\ref{eq:kappa-1}-\ref{eq:lambda3}) become
\begin{eqnarray}
\kappa_{cD} &=& \frac{1}{3}C(T) v^2 \overline{\tau_c} \nonumber \\
\lambda_{1D} &=& \frac{1}{3\hbar}C(T) \overline{\tau_c} \nonumber \\
\lambda_{2D} &=& \frac{1}{3\hbar}C(T) \overline{\tau_c /\tau_N} \nonumber \\
\lambda_{3D} &=& \frac{1}{3\hbar^2 v^2}C(T) \overline{\tau_c /\tau_U} 
\label{eq:lambdaD}
\end{eqnarray}
Then my result for the Callaway model in Debye approximation is
\begin{equation}
\kappa_{C}=\kappa_{\rm RTA}\left(1+\frac{\overline{\tau_c(\omega,T)/\tau_N(\omega,T)} }{\overline{\tau_c(\omega,T)/\tau_U(\omega,T)}}\right)
\label{eq:kappaD}
\end{equation}
Callaway's solution has
an extra factor $1/\tau_Q^N$ in both $\lambda_{1}$ and $\lambda_{3}$.  Then Eq.(\ref{eq:kappaD}) is
replaced by 
\begin{equation}
\kappa_{C}^\ast=\kappa_{\rm RTA}\left(1+\frac{\overline{\tau_c(\omega,T)/\tau_N(\omega,T)} ^2}
{\overline{\tau_c(\omega,T)} \ \ \overline{\tau_c(\omega,T)/\tau_U(\omega,T)\tau_N(\omega,T)}}\right)
\label{eq:kappaDold}
\end{equation}
The notations $\kappa_C$ and $\kappa_C^\ast$ denote the present (new and corrected) solution
of Callaway's model in Debye approximation, and the original (old and uncorrected) solution.

\begin{figure}[tbp]
\centering
\includegraphics[width=9.5cm]{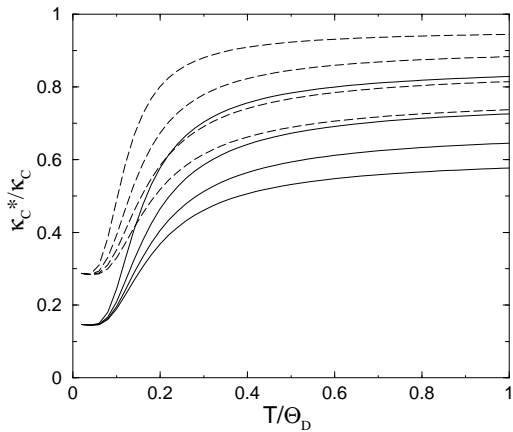}
\caption{Ratio of Callaway's old solution ($\kappa_C^\ast$) to the new solution ($\kappa_C$) of the Callaway model
in Debye approximation, with quadratic behavior $1/\tau_N(\omega) = \gamma_N(\omega/\omega_D)^2$.  
The dashed curves use the Herring quadratic behavior also for $U$, with ratio $\tau_N(\omega)/\tau_U(\omega)=
\gamma_U/\gamma_N=g$ and
$g=A\exp(-\Theta_D/3T)$.  The solid curves have $p_U=4$ or $\tau_N(\omega)/\tau_U(\omega)=g(\omega/\omega_D)^2$,
and the same form for $g$.  In both cases, the four curves, from lowest to highest,
are for values of $A$ set to 1, 2, 4, and 10. }
\label{fig:edelta}
\end{figure}

Finally, it is necessary to choose power laws, $p_N$ and $p_U$ for $N$ and $U$ scattering rates. 
Following Herring\cite{Herring}, the $N$-processes are 
assumed to have quadratic $\omega$ dependence, $p_N=2$.  This has been confirmed in recent numerical 
calculations\cite{Ward,Esfarjani}.  Herring also suggested quadratic behavior $p_U=2$ for the $\omega$ dependence
of $U$ processes.  However, numerical calculations have been recently fit to larger powers, $p_U=4$
(ref. \onlinecite{Ward}) and $p_U=3$ (ref. \onlinecite{Esfarjani}).  Results for two cases, $p_U=2$ (Herring)
and $p_U=4$ (a possible alternative) are shown in Fig.\ref{fig:edelta}.

Callaway's solution $\kappa_C^\ast$ underestimates the suppression of $N$-scattering, and thus underestimates
the thermal conductivity.  I believe that $\kappa_C$, the larger solution, is the true solution of Callaway's model.
In the low $T$ limit, integrals can be done analytically.  In Herring's case ($p_N=p_U=2$), the ratio 
$\kappa_C^\ast/\kappa_C\rightarrow 7/25$, and in the case $p_N=2$, $p_U=4$, the ratio becomes
 $\kappa_C^\ast/\kappa_C\rightarrow 1/7$.  At higher $T$, as $U$ scattering increases
to the level of $N$ scattering, the difference between Callaway's old solution
and the present new one is smaller.

The new solution simplifies in the case $p_N=p_U$, as, for example, in Herring's case where
both are 2.  The complexities in Eq.(\ref{eq:kappaD}) cancel, leaving 
$\kappa_C\rightarrow \kappa_{\rm RTA}
\times(1+\gamma_N/\gamma_U)$.  This is true at all $T$, leaving the simple answer
$\kappa_C=(1/3)Cv^2\overline{\tau_U(T,\omega)}$;  $N$-scattering drops out completely.  This is definitely
not true of Callaway's solution.
For more realistic models, for example, with $p_U=4$, frequency integrals are more complicated,
and $N$ scattering does not completely disappear.  In fact, since $U$ scattering has apparently
an $\omega^3$ or $\omega^4$ limiting behavior, and impurity scattering has the
Rayleigh form $\gamma_{\rm imp}\propto \omega^4$, the quadratic behavior of the $N$ term
is the only thing that prevents a low-$\omega$ divergence in the integrals.  A low-$\omega$
divergence is not completely unphysical.  At low enough $T$, the only vibrations thermally excited
are sure to have mean free paths longer than sample size.  Whether or not their contribution to the
integrals for $\kappa$ converge, they in fact do not contribute currents governed by temperature
gradients.  Instead, they give ballistic currents determined by the difference available heat
({\it i.e.} $T^4$) in the baths at the two ends.  The integrals have to be cut off at some $\omega_{\rm min}$
whose value depends on sample size.

\begin{figure}[tbp]
\centering
\includegraphics[width=9.5cm]{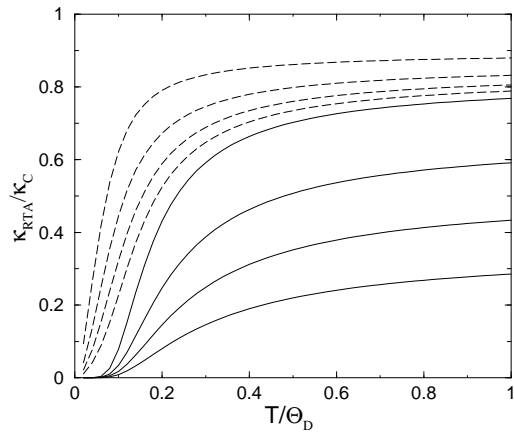}
\caption{Ratio of relaxation-time approximation ($\kappa_{\rm RTA}$) to full solution $\kappa_C$ for 
the Callaway model in Debye approximation 
with quadratic $\omega$ dependence for $1/\tau_N$ and quartic $\omega$ dependence
for $1/\tau_U$.  The four solid curves use the same parameters as the solid curves of Fig.\ref{fig:edelta}.  The
dashed curves have $A=10$ (as does the top solid curve), but also include Rayleigh-type impurity scattering
($1/\tau_{\rm imp}=\gamma_{\rm imp}(\omega/\omega_D)^4$), a momentum-non-conserving event
which adds to $1/\tau_U$ without the low-$T$ thermal suppression.  The strength of the impurity
term is $\gamma_{\rm imp}/\gamma_N=1, \ 2, \ 4, \ 10$ (from the lowest dashed curve to the highest.)}
\label{fig:ddelta}
\end{figure}

Fig.\ref{fig:ddelta} shows some of the same results as Fig.\ref{fig:edelta}, except in a different ratio, comparing with
the relaxation-time approximation rather than the original Callaway approximation.  Also shown is the
effect of including large amounts of momentum-non-conserving impurity scattering.  Enhancement
of thermal conductivity is still quite large, since $N$ scattering dominates $1/\tau_{\rm QP}$ at low $T$.
However, the enhancement is smaller because $1/\tau_{\rm imp}$ exceeds $1/\tau_U$ at lower $T$.

\section{Callaway's model and reality}

 Computation is now advanced enough to give a final answer to the question, how realistic is Callaway's model?
There are good algorithms for accurate construction and solution of the PBE, at least at $T$ not too low (where mesh size
becomes a problem because only small $\vec{Q}$ phonons are excited.)  
An iterative solution of the PBE begins with a first iteration which is the RTA.  This requires
 full computation of the quasiparticle
 relaxation rate, $1/\tau_Q^{\rm QP}$.  It is straightforward in principle to separate this into $N$ and $U$
 parts.  These could be used to obtain the Callaway model solution from  $\lambda_1, \ \lambda_2, \ 
  {\rm and} \ \lambda_3$ of Eqs.(\ref{eq:lambda1}-\ref{eq:lambda3}).  If a full interative solution
  of the PBE is then completed, it would be interesting to compare with the Callaway solution.

In the Herring version of the Debye approximation to the Callaway model, with anharmonic phonon scattering
rates going like $1/\tau_Q \propto \omega_Q^2$ for both $N$ and $U$, the answer is that $N$ processes drop out, and
the Umklapp scattering determines the conductivity.  This is a nice, but oversimple result.
Once the model gets more complex, with
multiple momentum relaxing processes with differing $\omega$-dependences, $N$ processes no longer drop out completely,
but cannot alone relax the distribution $N_Q$ to the zero-current distribution.  For models more realistic than Debye,
N-processes can relax the current toward a value $\hbar <v^2> \vec{P}/\Omega$, but still cannot relax the current
completely.

There has been a practice of computing only $\kappa_{\rm RTA}$, sometimes with the
claim that the Boltzmann equation has thus been solved.  Two arguments may seem to support this.  
First, Callaway used his solution
to fit quite accurately the measured $\kappa(T)$ of Ge.  He did not find much enhancement beyond RTA from the  
reduced role of N processes.  This must be, at least partly, an artifact of the inaccuracy
of his solution.  Thus Callaway's work seems to approximately
support the RTA, but the support cannot be taken seriously.  The second
argument is that the corrections are not often as big in complex materials as they are in the Debye
model shown in Figs. \ref{fig:edelta} and \ref{fig:ddelta}).

Several converged iterative solutions of the PBE have been reported that include some discussion
of the departure of the full solution from the RTA.
Ward {\it et al.} \cite{Ward2} in Fig. 2 show an 80\% increase in $\kappa(172{\rm K})$ for
diamond, by converging the PBE rather than using RTA.  The increase lowers to  30\% at 1200K.
However, in other systems, the error is often not so large.  For example,
Fig. 3 of Chernatynskiy {\it et al.} \cite{Phillpot} shows that for MgO, the effect on $\kappa(T=300{\rm K})$ is only a 7\%
increase, and the effect is smaller in UO$_2$.   In SrTiO$_3$, where the effect is tiny at 250K, it is 7\% at 50K  and a
42\% increase at 20K \cite{Alex}.  The message is, that for materials with complex phonon spectra, U processes are
only thermally suppressed at quite low $T$.  This is why the inability of N processes alone to degrade heat current
does not show up except at low $T$ and with pure samples (including isotopic purity).  
The recent prediction of large $\kappa$ in BAs \cite{Lindsay} is related to
weakness of U processes (and the resulting ineffectiveness of N processes)
caused by phonon dispersion that is quite simple and also unusual.  It would be interesting
to ask whether the Callaway model has decent predictive power in this case.  Even if not predictive,
the Callaway model has given needed insight, and should continue to do so.  Therefore,
the corrected (and simpler) solution of this model found in this paper should have some value.

{\bf Acknowledgements} \ \
 I am grateful to Tao Sun for discussions that motivated this work.  I thank D. Broido, A. Chernatynskiy,
K. Esfarjani,  and P. Schelling,  for helpful correspondence.  
The work was supported in part by DOE grant number DE-FG02-08ER46550.


\begin{thebibliography}{99}
\bibitem{footnote} The best choice is presumed to be when $1/\tau_Q^c=1/\tau_Q^{\rm QP}$.  Callaway's notation
$1/\tau_Q^c$ allows other models, presumably less accurate.
\end{thebibliography}
\end{document}